\documentstyle[aps,prl,multicol,epsf]{revtex}
\begin{document}
\draft
\title{One--parameter Superscaling at the Metal--Insulator
    Transition in Three Dimensions}
\author{Imre Varga$^{a,b}$, Etienne Hofstetter$^c$, and J\'anos
    Pipek$^a$}
\address{$^a$Elm\'eleti Fizika Tansz\'ek, Budapesti M\H uszaki Egyetem,
    H-1521 Budapest, Hungary}
\address{$^b$Fachbereich Physik, Philipps--Universit\"at Marburg,
    Renthof 6, D-35032 Marburg an der Lahn, Germany}
\address{$^c$Blackett Laboratory, Imperial College, London SW7 2BZ,
    United Kingdom}
\date{\today}
\maketitle
\begin{abstract}
    Based on the spectral statistics obtained in numerical simulations
    on three dimensional disordered systems within the tight--binding
    approximation, a new superuniversal scaling relation is presented
    that allows us to collapse data for the orthogonal, unitary and 
    symplectic symmetry ($\beta=1,2,4$) onto a single scaling curve. 
    This relation provides a strong evidence for
    one--parameter scaling existing in these systems which exhibit
    a second order phase transition. As a result a possible
    one-parameter family of spacing distribution functions, $P_g(s)$,
    is given for each symmetry class $\beta$, where $g$ is the dimensionless
    conductance.
\end{abstract}
\begin{multicols}{2}
\narrowtext
\newpage
    The study of critical phenomena is an important subject because of
    the rich variety of systems exhibiting a second order phase
    transition \cite{Bin}. By a second order transition we mean a
    continuous transition between two regimes with the the correlation 
    length diverging at the transition point. The description of
    such phenomenon leads to the introduction of very important
    concepts such as scaling, renormalization group and universality
    classes. These reflect the fact that the phase transition does
    not depend on the details of the system but only on some general
    symmetries as well as on the dimension of the system. A
    direct consequence is that different systems with different
    Hamiltonians may share the same critical exponents, describing the
    singularity of the phase transition, if the symmetry underlying 
    these systems is the same and therefore will belong to the
    same universality class. Other features, on the other hand, may be
    in common for different universality classes leaving the
    possibility to derive simple relations between these classes. Such
    a feature is scaling which is exploited in order to find the
    position of the critical point and the value of the critical
    exponent. Even though scaling may be commonplace,
    the scaling function may be different for the different
    universality classes.
\par
    In this Letter we present a single one--parameter scaling relation which
    is common to several different universality classes. This relation 
    involves the spectral statistics of a three dimensional
    (3D) disordered system with additional degrees of freedom, e.g.,
    strong magnetic field and spin--orbit scattering. The choice of this
    system comes from the realization that it exhibits a
    metal--insulator transition (MIT) as a function of the disorder in the 
    thermodynamic limit~\cite{Ang1}. It is generally assumed that
    the critical behavior at the MIT can be classified in terms of three 
    different
    universality classes according to the symmetry of the system:
    orthogonal [with time reversal symmetry, O$(N)$], unitary [without
    time reversal symmetry, e.g. with a magnetic field, U$(N)$] and,
    symplectic [with spin orbit-coupling, Sp$(N)$]. One then expects
    different critical exponents related to the MIT for the three
    different universality classes.
\par
    Surprisingly, in spite of the apparent change of universality
    class, the same value of the critical exponent has been found,
    numerically, both in the presence and absence of a magnetic field 
    \cite{Kra1,Kra2}, as
    well as spin-orbit coupling \cite{To,Eh1}. Moreover, Ohtsuki
    {\it et al.} recently showed \cite{Tom2} that the anomalous
    diffusion exponent and also the fractal dimension $D(2)$ seem to
    coincide at the MIT for O$(N)$, U$(N)$ or Sp$(N)$, in agreement
    with these results. It was recently proposed~\cite{Eh1,Eh3} that a
    natural way to understand these coincidences would be to invoke
    the spontaneous breaking of the symmetry right at the MIT.
    However, in a recent paper~\cite{ToS}, numerical evidence has been
    presented suggesting a small difference between the scaling
    properties of orthogonal and unitary systems.
\par
    The problem is therefore far from being solved and we wish to present new
    evidence concerning how the different universality classes are linked
    together. This indication gives a nontrivial hint about the way in which
    the symmetry
    parameter enters into the scaling function valid for each
    individual universality class. We also present a possible
    one-parameter family of spacing distribution functions, $P(s)$, 
    for each universality class.
\par
    A convenient way to study the MIT is to resort to random matrix
    theory (RMT) and energy level statistics (ELS)~\cite{Shkl,PhR,BA}. 
    In RMT the
    statistics of the energy spectrum are generally described by three
    different ensembles, Gaussian orthogonal (GOE), unitary (GUE), and
    symplectic (GSE) depending upon the symmetries mentioned
    above. Recently it has been shown~\cite{To,Eh1,Eh3,Shkl,Eh4,Isa}
    that in addition to the two expected statistics, namely either GOE, GUE 
    or GSE for the metallic regime and the Poisson ensemble (PE) for the
    insulating regime, there is a third statistics, called the critical
    ensemble (CE), which occurs only exactly {\it at} the critical point.
\par
    In order to investigate the MIT we consider the following
    tight--binding Hamiltonian~\cite{Ang1}
\begin{equation}
    H=\sum_{n} \epsilon_{n} |n><n| + \sum_{n,m} V_{n,m} |n><m|
\label{ham}
\end{equation}
    with
\begin{equation}
    V_{n,m}=\left \{ \begin{array}{ll}
      V \;\; & \;\;{\rm orthogonal} \\
      V \exp(i\theta_{n,m}) & \;\;{\rm unitary} \\
      V \exp(i \mbox{\boldmath $\theta$}_{n,m}) & \;\;{\rm symplectic}
                     \end{array}\right.
\end{equation}
    where the sites $n$ are distributed regularly in 3D space, e.g., on
    a simple cubic lattice. Only nearest neighbor interactions are 
    considered. The phase $\theta_{n,m}$ is a scalar related to
    the magnetic field \cite{Eh3} and $\mbox{\boldmath
    $\theta$}_{n,m}$ is a $2\times 2$ matrix \cite{Eh1}. The site
    energy $\epsilon_{n}$ is described by a stochastic variable. In
    the present investigation we use a box distribution with variance
    $W^2/12$. The parameter $W$ describes the disorder strength and is 
    the critical parameter.
\par
    Based on the above Hamiltonian, the MIT is studied by the ELS
    method, i.e., via the fluctuations of the energy spectrum
    \cite{Eh1,Shkl}. Starting from Eq. (\ref{ham}) the energy spectrum was
    computed by means of the Lanczos algorithm for systems of size $L
    \times L\times L$ with $L=13,\;15,\;17,\;19 \;{\rm and}\; 21$ and
    $W$ ranging from 3 to 100 averaging over different
    realizations of the disorder. After unfolding the spectra
    obtained, the fluctuations can be appropriately described by means
    of the spacing distribution $P(s)$~\cite{PhR}. This distribution 
    measures the level repulsion and is normalized as is its first moment:
    $\mu_1=\langle s\rangle =1$.
\par
    In order to characterize the shape of $P(s)$, we first calculate
    shape descriptive parameters which continuously change as we vary 
    external parameters, e.g., the system size $L$ or disorder $W$:
\begin{equation}
    q=\mu_2^{-1}\quad \hbox{and}\quad S_{str}=\mu_S +\ln \mu_2,
\end{equation}
    where  $\mu_2=\langle s^{2}\rangle$ is the second moment of
    $P(s)$, while $\mu_S=-\langle s \ln s \rangle$. These
    quantities were first introduced to describe the
    spatial--localization properties of general lattice distributions
    \cite{Imr1} and then used for the shape analysis of $P(s)$ around
    the MIT \cite{Imr2}. It is interesting to note that in contrast to
    previous methods which used only part of the information contained in
    $P(s)$ \cite{Shkl,Eh4} we consider here the entire distribution
    obtained numerically. Parameter $q$ is a well--know quantity in
    probability theory that describes the peakedness of a distribution
    function. For example, for $P(s)=\delta (s-1)$ we have $q=1$. The
    parameter $S_{str}$ is
    called the structural entropy for reasons described elsewhere
    \cite{Imr1}. These parameters describe not only the bulk features of 
    $P(s)$, but also they are sensitive to the numerical upper cutoff of 
    the support of $P(s)$.
\par
    In order to describe and compare the different universality classes
    within the same method we perform a linear rescaling as
\begin{mathletters}
\label{qstilde}
\begin{eqnarray}
    \label{qtilde}
    -\ln (q)&\to &
    \frac{-\ln (q)+\ln (q_W)}{-\ln (q_P)+\ln (q_W)}=\tilde Q
    \\ \label{stilde}
    S_{str}&\to &
    \frac{S_{str}-S_W}{S_P-S_W}=\tilde S
\end{eqnarray}
\end{mathletters}
\noindent
    where index $_{P}$ refers to the PE and $_{W}$ to the
    Wigner--surmise representing the GOE, GUE or GSE respectively. 
    The choice of such a rescaling defined 
    in Eqs.~(\ref{qstilde}) maps the variables $\tilde S$
    and $\tilde Q$ onto the $[0,1]$ interval, with $\tilde S=\tilde
    Q=0(1)$ belonging to the RMT (PE) limit. Furthermore, in
    Eq.~(\ref{qtilde}) it is more natural to use $-\ln(q)$ instead of
    $q$ since, similarly to $S_{str}$, it is connected to differences
    of R\'enyi--entropies~\cite{renyi}.
\begin{figure}[tbh]
\epsfxsize=3in
\epsfysize=3in
\epsffile{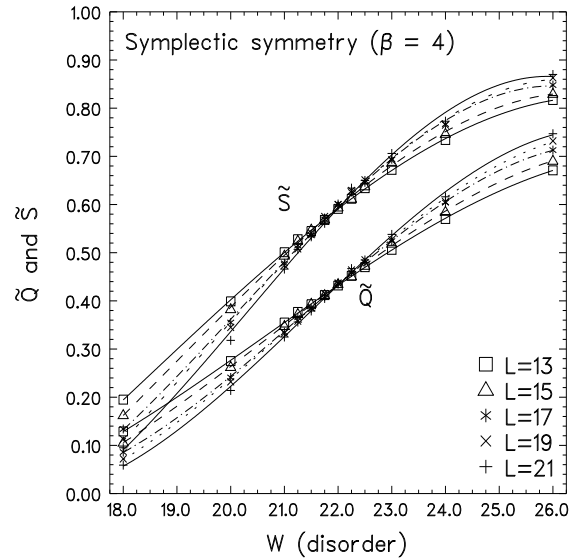}
\vspace{3mm}
\caption[beta=4]{\label{b4} $\tilde Q(L,W)$ and $\tilde S(L,W)$ for
    the case of symplectic symmetry. Continuous curves are polynomial
    fits.}
\end{figure}
    As an illustration of the behavior of these parameters, in Fig.~\ref{b4}
    we report the results for $\tilde Q(L,W)$ and $\tilde S(L,W)$ for the case
    of spin-orbit coupling ($\beta=4$). We can see that the data
    depend on the size of the system except at the critical point
    $W_{c}$ where $P(s)$ is scale invariant. This is due to the fact
    that the MIT is a second order transition and that finite-size
    scaling laws apply close to the transition~\cite{Fish}. These
    properties were already used with success to describe the MIT
    \cite{Shkl,Eh4,Eh5}. In particular it was shown that such
    quantities have a finite size scaling behavior and can be written
    as
\begin{equation}
    q(L,W)=f(L/\xi_{\infty});\,\,S_{str}(L,W)=h(L/\xi_{\infty})
\label{sq}
\end{equation}
    with correlation length $\xi_{\infty}(W)\sim |W-W_{c}|^{-\nu}$, 
    and the critical exponent $\nu$. The functions $f(x)$ and $h(x)$ are
    universal in the sense that they do not depend on the
    details of the systems - just on the general symmetries - and
    therefore they directly reflect the universality class of the system.
\par
    From Eq.~(\ref{sq}) we can see that, because of the scaling
    behavior of $q(L,W)$ and $S_{str}(L,W)$, if we plot $S_{str}$ as a
    function of $q$ we can see the similarities and also the
    differences between the universality classes. The same is true for
    the rescaled parameters $\tilde{Q}$ and $\tilde{S}$.
\par
\begin{figure}[tbh]
\epsfxsize=3in
\epsfysize=3in
\epsffile{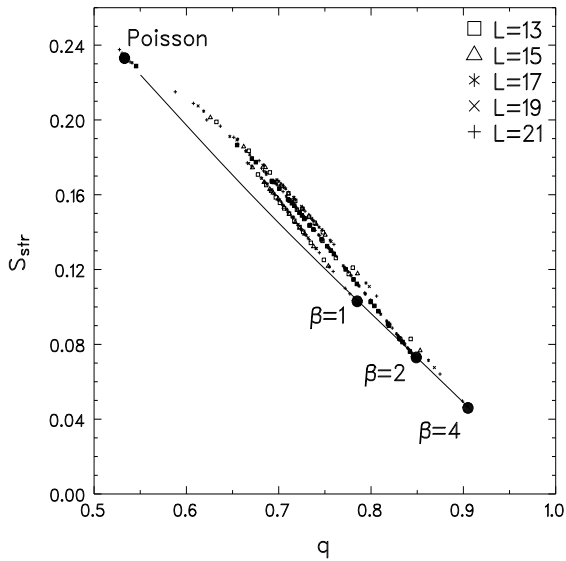}
\vspace{3mm}
\caption[all beta 1]{\label{qS} $S_{str}$ as a function of $q$
    for all the symmetry classes. (All data are presented in the full
    range of disorder.) The solid curve is obtained from a simple
    interpolating $P(s)$ due to Izrailev~\protect\cite{izr}. The RMT
    and Poisson distributions appear as solid circles.}
\end{figure}
\begin{figure}[tbh]
\epsfxsize=3in
\epsfysize=3in
\epsffile{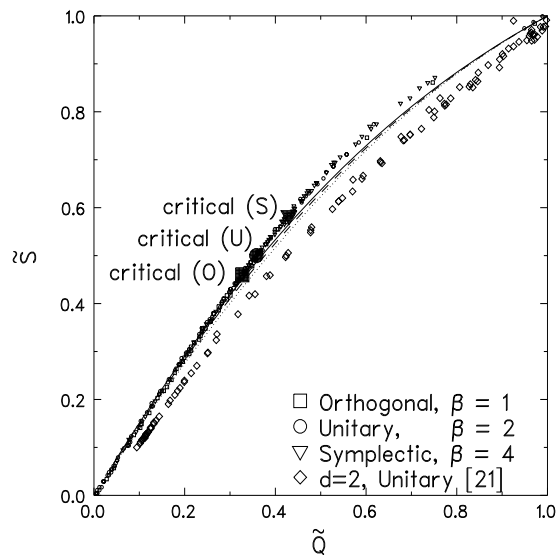}
\vspace{3mm}
\caption[all beta]{\label{b-all} $\tilde S(L,W)$ as a function of
    $\tilde Q(L,W)$ for all the symmetry classes.
    The solid symbols represent the positions of the critical points.
    The continuous (solid, dashed, dotted) curves are our analytical
    estimates, see text for details. For comparison the results obtained
    for the network mode~\protect\cite{Var} of the quantum Hall effect 
    ($d=2$, $\beta=2$) are also presented.}
\end{figure}

    Indeed, Fig.~\ref{qS} shows clear differences between the
    orthogonal, unitary and symplectic cases, although all the data
    fall onto special curves irrespective of $W$ and $L$ for each
    case. This figure allows us to determine the scaling relations
    for $\beta=1,2$ and $4$ without having to derive
    $f(L/\xi_{\infty})$, $h(L/\xi_{\infty})$ and $\xi_{\infty}(W)$
    which are not easy to obtain numerically due to their
    singularities at the critical point.
    Using now the rescaling defined above in Eqs.~(\ref{qstilde}), we
    plot $\tilde S(L,W)$ as a function of $\tilde Q(L,W)$. The results
    are shown in Fig.~\ref{b-all}. We see that all the data scale
    nicely onto the {\it same} curve indicating the presence of a 
    one--parameter superscaling function. The position of the MIT moves
    along the {\it same} curve, for $\beta=1,2$ and $4$, as a function
    of the critical disorder $W_c$ which can be changed by the
    magnetic field and spin-orbit scattering rate as well as the type
    of potential scattering. This new superscaling relation is very
    interesting and of importance in shedding new light on the MIT in 3D
    systems. New results~\cite{Var} indicate that the data for 
    $\beta=2$ in 2D scale onto a different curve (see Fig.~\ref{b-all}).
    This point is important because
    it implies that superscaling is not a mere consequence of the
    universality of level repulsion but something more subtle.
\par
    Next we will show that the observed relation $\tilde S(\tilde Q)$
    presented in Fig.~\ref{b-all} can be understood with the introduction 
    of the dimensionless conductance as a scaling variable. We have
    found that the constant shifts in (\ref{qstilde}) for both
    terms $-\ln (q)+\ln (q_W)$ and $S_{str}-S_W$,
    correspond to a convolution of different distributions \cite{PV2}:
\begin{equation}
    {\cal P}_{g,\beta}(e^x)=
    \int_{-\infty}^{\infty }{\cal Q}_{g,\beta}(e^{x-y})
    {\cal W}_{\beta}(e^y)dy
\label{conv}
\end{equation}
    with $e^x\equiv s$. In this case~\cite{PV2} 
    $-\ln(q_{\cal P})=-\ln(q_{\cal Q})-\ln(q_{\cal W})$ and also
    $S_{str}^{\cal P}=S_{str}^{\cal Q}+S_{str}^{\cal W}$. 
\par
    In Eq.~(\ref{conv}), ${\cal P}_{g,\beta}(s)$ is the numerically 
    obtained spacing
    distribution for different symmetry classes parametrized by the
    dimensionless conductance $g$, which ranges from zero to infinity as
    $L$ and $W$ changes as well, and ${\cal W}_{\beta}(t)$ is the RMT
    limit for $g\to\infty$ represented by, e.g., the Wigner--surmise.
    This rescaling provides us with a method to study what
    is {\it beyond} the universal level repulsion present in finite systems
    in the full range of disorder.
    The parameters $-\ln(q)$ and $S_{str}$ of ${\cal P}_{g,\beta}(s)$ 
    give different curves [see Fig.~(\ref{qS})], while after rescaling
    $\tilde Q$ and $\tilde S$ give the {\it same} curve 
    [see Fig.~(\ref{b-all})]. This is what is meant by the {\it superscaling} 
    relation as can be seen in Fig.~\ref{b-all}. Scaling in this 
    context refers to the appearance of $g$. We will show that the
    parameters $\tilde Q$ and $\tilde S$ of the function 
    ${\cal Q}_{g,\beta}(s)$ appearing in Eq.~(\ref{conv}) can account
    for the major part of the numerically observed relation.
\par
    In what follows we give an approximate formula for
    ${\cal P}_{g,\beta}(s)$ based on analytical calculations and a
    phenomenological assumption. First, we point out that
    Eq.~(\ref{conv}) can be solved exactly for the extreme cases of a
    perfect metal ($g\to\infty$) and perfect insulator ($g\to 0$). In
    the former case the left hand side should equal the Wigner surmise
    and it is easy to show that such a convolution will hold if the
    ${\cal Q}_{g,\beta}(s)$ function as $g\to\infty$ approaches a
    Dirac--delta function, $\delta(s-1)$. As for the perfect insulator we have
    to find ${\cal Q}_{0,\beta}$ so that the left hand side in each
    case equals $P(s)=\exp(-s)$. The solution for these problems,
    introducing the notation $R(s)\equiv {\cal Q}_{0,\beta}(s)$,
    is~\cite{PV2}
\begin{equation}
    R(y)=a\left \{
    \begin{array}{ll}
        e^{-y^2}                 & \;\;\beta=1 \\
        {\rm erfc}(y)                 & \;\;\beta=2 \\
        (2y^2+1){\rm erfc}(y)-\frac{2y}{\sqrt{\pi}}e^{-y^2}
                                 & \;\;\beta=4
    \end{array}\right.
\label{wp}
\end{equation}
\noindent
    where $y=bs$, $a=2/\pi$, $\pi/4$, and $9\pi/64$, and
    $b=1/\sqrt{\pi}$, $\sqrt{\pi}/4$, and $3\sqrt{\pi}/16$ for
    $\beta=1$, $2$ and $4$, respectively.
\par
    These solutions are spacing distributions themselves since their zeroth
    and first moments are normalized to unity. The interpolating formula is
    introduced based on the most simple assumption
\begin{equation}
    {\cal Q}_g(s)= a_gs^gR(b_gs)
\label{interp}
\end{equation}
\noindent
    Parameters $a_g$ and $b_g$ are determined from the normalization
    conditions $\langle 1\rangle=\langle s\rangle=1$ for each $\beta$.
    These interpolating distributions behave in the limit $g=0$ and
    $g\to\infty$ appropriately as defined above. The continuous curves
    in Fig.~\ref{b-all} show that the rescaled R\'enyi--entropies of
    ${\cal Q}_g(s)$ [Eq.~(\ref{interp})]
    indeed reproduce the results of the numerical experiments. Solid,
    dashed, and dotted lines stand for $\beta=1$, $2$, and $4$,
    respectively. However, we see that the analytical curves do {\it
    not} fall onto the same curve. This discrepancy may be due
    to the simplicity of the approximation in (\ref{interp}) and also
    because of the presence of a maximal spacing, i.e. a cut--off in
    both the numerical histogram and consequently in the analytical
    curves. The analytical curves without the upper cut--off (not 
    presented here) fall on top of each other within the linewidth precision. 
\par
    Finally, Fig.~\ref{b-all} allows us to give an estimate of the
    critical conductance $g^*$. The best fits to the numerical
    histograms give $g^*=1.58$, 1.46, and 1.34 for $\beta=1$, 2, and 4,
    respectively.
\par
    In conclusion, we have presented evidence for a new superscaling
    relation characterizing the MIT in 3D disordered systems with
    different additional degrees of freedom, i.e., in different
    universality classes. Such a relation gives a hint for the
    derivation of the symmetry dependence of the scaling function.
    We have also given an approximate analytical formulation of the
    spacing distribution where the symmetry parameter $\beta$ and the
    scaling variable $g$ enter in a very clear way. The estimates of
    the critical conductance on the other hand show differences for
    the position of the MIT. This result is complementary to the fact
    that the critical exponent $\nu$ obtained numerically in the three
    cases is the same~\cite{Kra1,Kra2,To,Eh1}.
\par
    We have to note that in some recent experiments providing the same
    value of the critical exponent~\cite{Loe,Hor}, as well as the
    absence of influence of the magnetic field~\cite{Hor} and the
    spin--orbit coupling~\cite{Loe} at the MIT, show the possibility
    that the superscaling relation presented in this Letter could be
    verified experimentally. 
\par
    The method presented in this Letter can be useful in the analysis
    of other phase transitions as well.
\par
    {\bf Acknowledgment:} We would like to thank M. Blencowe for reading 
    the manuscript. Financial support from Orsz\'agos
    Tudom\'anyos Kutat\'asi Alap (OTKA), Grant Nos. T029813, T024136
    and F024135 are gratefully acknowledged.

\end{multicols}
\end{document}